\documentclass{aa} 

\usepackage{graphicx}
\usepackage{natbib}         
\usepackage{txfonts}
\usepackage{xcolor}
\usepackage{hyperref}
\usepackage{dblfloatfix}

\begin{document} 
   \title{Invariant manifolds in barred galaxy simulations}

   \subtitle{II. Quantitative evidence of manifold-trapping in spiral arm formation}

   \author{T. Soler-Terricabras
          \inst{1}\fnmsep\inst{2}\fnmsep\inst{3},
          M. Romero-Gómez\inst{1}\fnmsep\inst{2}\fnmsep\inst{3}
          \and
          S. Roca-Fàbrega\inst{4}
          }
    \titlerunning{Invariant manifolds in barred galaxy simulations -- II}
    \authorrunning{Soler-Terricabras et al.}

   \institute{Departament de Física Quàntica i Astrofísica (FQA), Universitat de Barcelona (UB), c. Martí i Franquès, 1, 08028 Barcelona, Spain\\
              \email{tsoler@fqa.ub.edu}
        \and
             Institut de Ciències del Cosmos (ICCUB), Universitat de Barcelona (UB), c. Martí i Franquès, 1, 08028 Barcelona, Spain
        \and
             Institut d’Estudis Espacials de Catalunya (IEEC), Edifici RDIT, Campus UPC, 08860 Castelldefels (Barcelona), Spain
        \and 
             Lund Observatory, Division of Astrophysics, Department of Physics, Lund University, SE-221 00 Lund, Sweden
             }

   \date{Received April 2, 2026; accepted May 26, 2026}

  \abstract
   {Understanding the nature of spiral arms in disc galaxies remains an open problem. Invariant manifolds associated with bars provide a framework in which particles with specific kinematic properties (manifold-compatible) are naturally able to support spiral structure, although their contribution to the observed total overdensities is yet to be quantified.}
   {The main goal of this work is to quantify, through a robust methodology, the contribution of invariant manifolds to the formation of spiral arms in a pure $N$-body simulation, setting up a machinery to perform similar tests in other and more complex simulations.}
   {We computed the invariant manifolds associated with the hyperbolic equilibrium points of the potential and quantified the fraction of particles whose motion is governed by these phase-space structures. We then compared the temporal evolution of this trapped fraction with the strength of the spiral arms, traced by the $A_2$ Fourier amplitude.}
   {We find a correlation between the fraction of trapped particles in the unstable exterior branches of the invariant manifolds and the strength of spiral arms. In particular, we determine that up to $50\%$ of all the particles located on the spiral arms region (and up to $90\%$ from the manifold-compatible population) are trapped by the manifolds, with oscillations of period $\sim 100$ Myr.}
   {Invariant manifolds provide a dynamically relevant framework for understanding the formation of spiral structure in pure $N$-body simulations of barred galaxies.
   We present the first quantitative evidence, based on a fully self-consistent $N$-body model, that a significant fraction of spiral-arm particles is governed by manifold-driven dynamics. 
   These particles act as seeds of overdensities that subsequently evolve into fully developed spiral arms through the delayed gravitational response of the disc to the self-gravity of the manifold-trapped material. The influence of the invariant manifolds remains non-negligible at all times, and phases of stronger spiral structure are associated with higher trapped fractions.
}

   \keywords{galaxies: spiral --
             galaxies: structure --
             galaxies: kinematics and dynamics --
             galaxies: formation --
             galaxies: evolution
               }

   \maketitle

\section{Introduction}\label{sec:introduction}

    \begin{figure*}[!ht]
       \centering
       \includegraphics[width=\hsize]{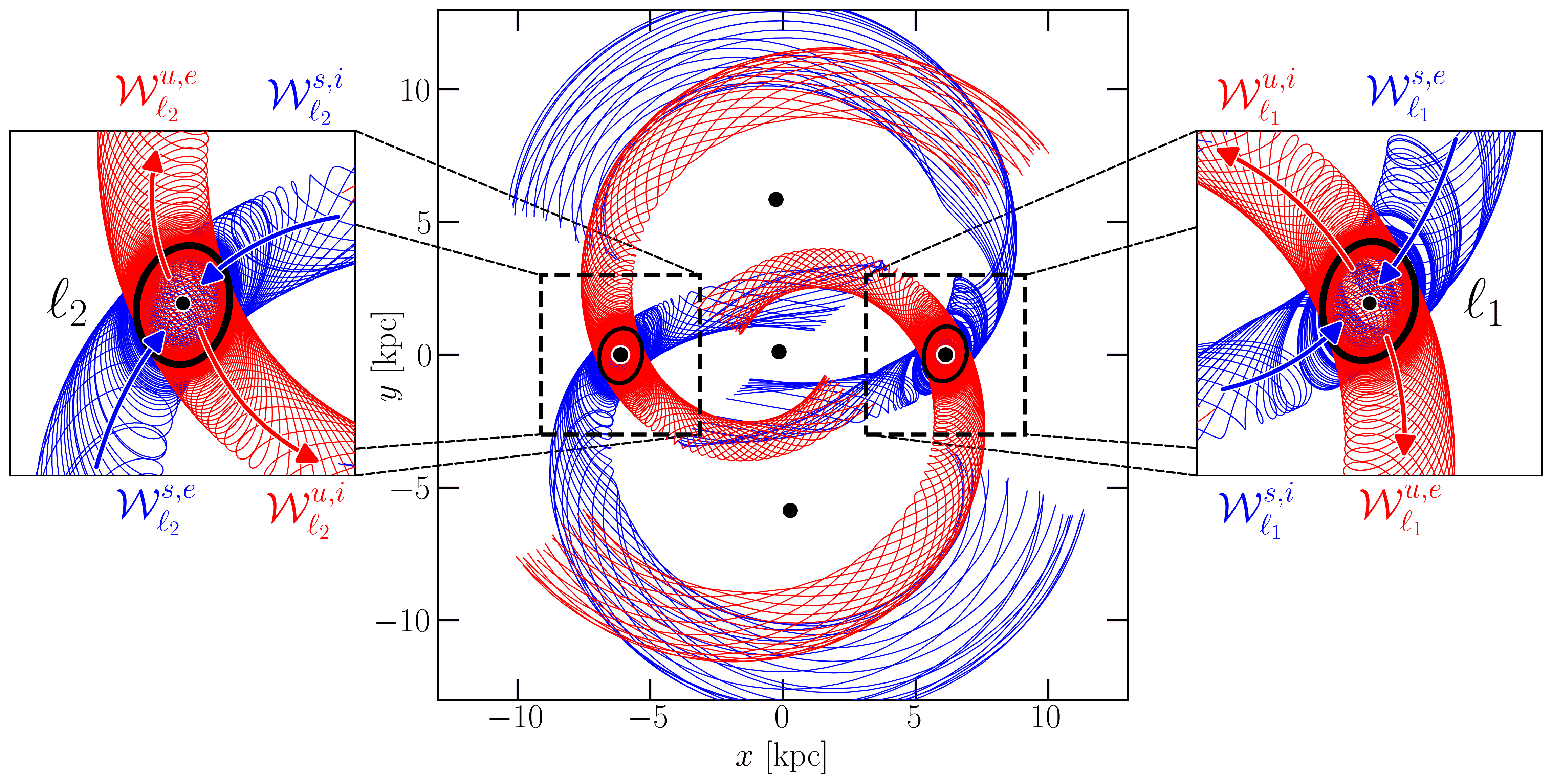}
            \caption{Schematic representation of the different branches of the invariant manifolds associated with the potential of a barred galaxy. The black dots mark the locations of the Lagrangian points, $L_i$, $i \in \{1,...,5\}$. Each of the black closed curves around $L_1$ and $L_2$ respectively represent the Lyapunov orbits, labelled as $\ell_1$ and $\ell_2$ by simplicity. The red curves represent the unstable branches of the invariant manifolds, $\mathcal{W}_{\ell_i}^u$, $i \in \{1,2\}$, which emanate from the Lyapunov orbits, while the blue curves display the stable branches, $\mathcal{W}_{\ell_i}^s$, $i \in \{1,2\}$, which asymptotically approach them. In the zoomed-in subplots, arrows have been overlaid on top of each manifold branch to illustrate the motion along them. For each branch, the superscripts $e$ and $i$ distinguish the exterior and interior branches, respectively, defined relative to the bar region.}
    
             \label{fig:B1_sketch}
    \end{figure*}

Bars constitute one of the most common large-scale features in disc galaxies: optical surveys identify bars in $30$–$50\%$ of nearby discs \citep{SellwoodWilkinson1993, Eskridge2000, Sheth2008, Masters2011, Lee2012}, rising to $50$–$70\%$ in the near-infrared \citep{Knapen2000, Eskridge2000, Whyte2002, Menendez-Delmestre2007, MarinovaJogee2007}.
Most barred galaxies also exhibit spiral structure \citep{Elmegreen1982, Buta2015}, and observations of grand-design spirals show that two-armed patterns dominate in barred systems  \citep{Elmegreen1982, Hart2016}, indicating that $m=2$ {Fourier} modes play a key role in shaping disc morphology.
Despite their ubiquity, the physical origin of spiral arms in barred galaxies remains unresolved.

Several mechanisms have been proposed to explain spiral structure in systems where the bar plays a central dynamical role. 
These include direct bar-driven responses \citep{KormendyNorman1979, TremaineWeinberg, Rautiainen1999, Rautiainen2000, Salo2010, GarmaOehmichen2021}, amplification through swing mechanisms \citep{GoldreichLyndenBell1965, JulianToomre1966, Toomre1981}, and recurrent or self-excited scenarios in which the bar acts as a long-lived source of non-axisymmetric forcing \citep{MassetTagger1997, Athanassoula2012, SellwoodCarlberg2014, SellwoodCarlberg2019, SellwoodCarlberg2021, SellwoodCarlberg2022}.
Historically, density-wave theory \citep{LinShu1964, LinShu1966, LinShu1969} has described spiral arms as long-lived, quasi-stationary patterns rotating with a constant pattern speed. 
However, numerical simulations have repeatedly challenged this picture \citep{SellwoodCarlberg1984, CarlbergFreedman1985}, instead revealing spiral arms as transient and recurrent features that break apart and reform over time \citep{Rautiainen1999, Rautiainen2000, Wada2011, Sellwood2011, Grand2012b, Roskar2012, RocaFabrega2013, Baba2013}. 
This tension between long-lived density waves and transient spirals lies at the heart of the lively debate about the true nature of spiral structure.

Adding to this ongoing debate, in the first paper of this series \citep[][hereafter Paper I]{TST2026a}, we proposed a material {density} wave scenario driven by invariant manifolds.
In this picture, unstable periodic orbits around the bar’s Lagrangian points $L_1$ and $L_2$ generate invariant manifolds that guide chaotic orbital flows away from the bar, forming {transient spiral arms populated by particles flowing along these phase-space tubes} \citep{MRG2006, MRG2007, Athanassoula2009a, Athanassoula2009b, Athanassoula2010}. 
Manifolds act as dynamical pathways that channel stars and gas into the outer disc, producing spiral features that continuously form, evolve, and dissolve as the bar-driven reservoir is replenished.
Fig. \ref{fig:B1_sketch} shows an schematic representation of the configuration of the invariant manifolds branches of a barred galaxy.

Although invariant manifolds have been explored in analytical, semi-analytical, {and self-consistent} models \citep{Voglis2006b, Tsoutsis2008, Athanassoula2012, Baba2015, Lokas2016, Efthymiopoulos2019, Efthymiopoulos2020, Zouloumi2024}, {providing detailed comparisons between particle distributions and invariant manifold structures,} their quantitative contribution to spiral structure in fully self-consistent $N$-body simulations of barred galaxies remains poorly constrained. 
In this work, we address this gap by {providing, for the first time, an explicit particle-based quantification of} the role of invariant manifolds in shaping spiral arms in a pure $N$-body simulation. 

In Paper I, we identified three dynamically distinct populations in the disc based on their Jacobi energies $E_J$, among which only the manifold-compatible\footnote{{As defined in Paper I of this series, manifold-compatible particles are those whose Jacobi energy is $E_{L_{1,2}}\leq E_J\leq E_\text{man}$, where $E_{L_{1,2}}$ is that of the saddle equilibrium points $L_1-L_2$, and $E_\text{man}$ is that of the invariant manifold, $E_\text{man}$. This population represents $\sim 30-40\%$ of the disc population.}} can trace the kinematic imprints of the spiral arms. 
{The kinematic signature of the low-energy population\footnote{{Low-energy particles are those whose Jacobi energy is $ E_J < E_{L_{1,2}}$, which represent the bulk of the disc and consistently account for $\sim 60\%$ of the disc population. See Paper I.}}, on the contrary, suggests that these particles respond to the spiral perturbation induced by the self-gravity of the material flowing along the invariant manifolds, thereby enhancing the density contrast of the arms.}
In the present paper, we develop a strategy to quantify the trapping of the manifold-compatible population within the unstable exterior branches of invariant manifolds\footnote{
{We denote by $\mathcal{W}^{u,e}_{\ell_i}$ the exterior (e) branch of the unstable (u) invariant manifold associated with the Lyapunov orbit $\ell_i$ around the equilibrium point $L_i$, for $i \in \{1,2\}$ (see Fig. \ref{fig:B1_sketch}).} Since these are the only branches of the invariant manifolds directly analysed in this work, we henceforth write $\mathcal{W}^{u}_{\ell_i} \equiv \mathcal{W}^{u,e}_{\ell_i}$, for $i \in \{1,2\}$, to simplify the notation.}, and compare the effectivity of the manifold-trapping to the evolution of spiral arm strength.
The method employed in this task is explained in Sec. \ref{sec:methodology}, with both a definition of the trapping criterion (Sec. \ref{sec:criterion}) and an assessment of its robustness (Sec. \ref{sec:validation}).
The results of this analysis are presented in Sec. \ref{sec:results}. 
Finally, the conclusions derived from the findings are summarised in Sec. \ref{sec:conclusions}.
   
\section{Methodology}\label{sec:methodology} 
    In this work, and as in Paper I, we analysed the B1 simulation of \citet{RocaFabrega2013}, a pure $N$-body model of an isolated barred galaxy, with $10^6$ stellar particles accounting for $M_{d}=5.0\times10^{10}\,M_\odot$.
    The 36 snapshots analysed in this work have been selected to ensure the presence of a well-defined non-axisymmetric structure: a strong bar with Fourier amplitude $A_2\gtrsim0.3-0.5$ , and two prominent, grand-design, bisymmetric spiral arms with $A_2\gtrsim0.1-0.3$ --see Fig. 1a in Paper I.
    These values are well above the commonly adopted thresholds for clearly defined bars \citep[$A_2\gtrsim0.2$, e.g.][]{Kraljic2012, Reddish2022} and bisymmetric spiral arms \citep[$A_2\gtrsim0.1$, e.g.][]{RocaFabrega2013, Quinn2026}.
    The sequence spans 561\,Myr, during which the bar completes two and a half revolutions, right after disc stabilisation and bar formation. 
    {In this particular set-up, the bar–spiral morphology corresponds to a configuration in which the spiral arms are connected to the ends of the bar at all times and co-rotate at their connection, sharing a common pattern speed, as reported in \citet{RocaFabrega2013} (see Fig. 4 therein). 
    This implies a strong dynamical coupling between the bar and the inner spiral structure, consistent with a manifold-driven or bar-driven spiral scenario. Beyond the bar–spiral interface, the spiral pattern speed decreases only mildly with radius.}
    
    The first step in this analysis was to construct the gravitational potential of each snapshot combining the snapshot-characterisation method proposed by \citet{Dehnen2022} and the potential construction tool provided by \texttt{AGAMA} \citep{Vasiliev2019}.
    See Paper I for further reference.
    As shown in Fig. 1b of Paper I, the bar pattern speed remains nearly constant, decreasing slightly from $30$ to $25 \, \text{km}\,\text{s}^{-1}\,\text{kpc}^{-1}$ in the analysed interval.
    We then proceeded with the computation of the equilibrium points and the subsequent construction of the invariant manifolds as described in \citet{MRG2006, MRG2007}. 
    In the subsequent analysis, we restricted the sample to particles close to the disc midplane ($|z|<400\,\mathrm{pc}$) and with small vertical velocities ($|v_z|<20\,\mathrm{km}\,\mathrm{s}^{-1}$), thereby focusing primarily on planar motion, as in Paper I.
    {This approximation provides a robust and appropriate framework for the present analysis: the simulated galaxy is isolated, and previous studies have demonstrated that the main invariant-manifold-driven orbital structures are largely preserved in the planar limit \citep[e.g.][]{OllePfenniger1998, KatsanikasPatsis2022, HarsoulaKatsanikas2025}.}

\subsection{Manifold-trapping criterion}\label{sec:criterion}

    {Let $\mathbf{\Omega_b}=\Omega_\text{b}\mathbf{e_z}$ denote the pattern speed of the bar, with $\Omega_\text{b}>0$ by convention (corresponding to counter-clockwise rotation), and let $(r,\phi)$ be the polar coordinates in the frame corotating with it.
    Then, the planar motion of particles in this rotating frame is governed by the Hamiltonian:}
    \begin{equation}\label{eq:jacobi}
        E_J\equiv\mathcal{H}(r,\phi,v_r,v_\phi)=\frac{1}{2}(v_r^2+v_\phi^2)+\Phi_\text{eff}(r,\phi)\,,
    \end{equation}
    {where $v_r$ and $v_\phi$ are radial and tangential velocities, and the effective gravitational potential, $\Phi_\text{eff}$, is defined as}
    \begin{equation}\label{eq:poteff}
        \Phi_\text{eff}(r,\phi) \equiv \Phi(r,\phi) - \frac{1}{2}\Omega_\text{b}^2\,r^2.
    \end{equation}
    {As shown in Eq. \ref{eq:jacobi}, the Hamiltonian depends on both position and velocity, $\mathcal{H}=\mathcal{H}(r,\phi,v_r,v_\phi)$.
    Therefore, the planar motion of a particle in the galaxy is determined by these four quantities: radius, azimuth, radial velocity and tangential velocity. 
    In this context,} we define a particle as \textit{trapped} by invariant manifolds when each of these four quantities remains within the bounds imposed by the manifolds.\\

    {Manifold-trapped particles must have Jacobi energies\footnote{In classical mechanics, when the Hamiltonian does not depend explicitly on time and the potential is independent of the velocities, the Hamiltonian coincides with the total energy of the system. Under these conditions, the Jacobi energy is therefore equivalent to the Hamiltonian of the system, and both quantities are conserved along the trajectories \citep{Arnold1989, Goldstein2002}.} in the range $E_{L_{1,2}} \leq E_J \leq E_\text{man}$ in order to be compatible with these phase-space structures.
    We refer to these population of particles as manifold-compatible.
    As discussed in Appendix A in Paper I, the Jacobi energy of a particle restricts its accessible phase-space regions. 
    Thus, selecting manifold-compatible particles already imposes a constraint on one of the four phase-space coordinates of the Hamiltonian.}

    {The projection of the invariant manifolds onto configuration space is not arbitrary, but is }intrinsically linked to the properties of the Lyapunov orbits around the saddle equilibrium points and the overall density distribution. 
    To characterise the spatial extent of the unstable exterior branches of the invariant manifolds, we apply constant-azimuth sections $\mathcal{S}$ to their configuration-space projection. 
    Specifically, we consider azimuthal angles from $\theta=10^\circ$ to $\theta=110^\circ$, in steps of $\Delta\theta=2^\circ$, measured from the positive $x$-axis.
    This azimuthal range is motivated both geometrically and dynamically. 
    From a geometric perspective, it corresponds to the region where the manifolds overlap with the bisymmetric, grand-design spiral arms. 
    From a dynamical point of view, the immediate vicinity of the equilibrium points $L_1$ and $L_2$ is dominated by chaotic motion \citep{Llibre1985, Xia1992, KaufmannContopoulos1996}, which prevents a clear and stable identification of particles following the invariant manifolds near the equilibrium region. 
    For this reason, the analysis begins at $\theta=10^\circ$, where trajectories start to move away from the strongly chaotic domain and a more consistent phase-space characterisation becomes possible.
    At the opposite end, beyond $\theta=110^\circ$, for the energy levels considered in the construction of the manifolds in this $N$-body simulation (see Paper I), the invariant manifolds gradually lose their well-defined structure beyond this azimuthal range \citep{Voglis2006b}, rendering further intersections less meaningful for the analysis. 
    Consequently, these azimuthal cuts allow for a systematic control of the manifold extent across different angles, and bound the second (out of four) degree of freedom.

    {The intersection of the manifold branches with each constant-azimuth section $\mathcal{S}$ yields} a closed curve onto the $r$-$v_r$ plane (Fig.~\ref{fig:B1_corba}a, inset panel), which we denote by $\mathcal{W}_{\ell_i}^{u,1}$, $i\in\{1,2\}$ \citep[as defined in][]{MRG2006, MRG2007}. 
    Each of these intersections establishes two extreme radial values, $r_{\rm min}$ and $r_{\rm max}$. 
    These values delimit the range of radial positions that are compatible with the manifold at each particular azimuth.
    Together, these radial bounds across all sampled azimuths constrain the third degree of freedom (out of four) that determines if a particle is trapped.

    The remaining degree of freedom must be constrained by one of the velocity components.
    In this work, we define the criterion by setting bounds on radial velocity, although the analysis is equivalent considering azimuthal velocity instead.
    As {shown} in Fig.~5 of \citet{MRG2006}, particles whose radial velocity lies on the curve $\mathcal{W}_{\ell_1}^{u,1}$ correspond to asymptotic orbits of the unstable branch of the invariant manifold, whereas particles whose radial velocities are enclosed within it correspond to orbits that flow along the manifold. 
    Conversely, particles whose radial velocities lie outside this curve in the $r$–$v_r$ plane are not trapped by the manifold dynamics.
    We therefore quantify the number of particles whose motion is guided by the manifolds by identifying those {manifold-compatible} particles that fall within the $\mathcal{W}_{\ell_1}^{u,1}$ curve in the $r$–$v_r$ plane, subject to the corresponding azimuthal and radial bounds. 
    {This strategy provides a constraint on the fourth and final degree of freedom, thereby enabling a precise identification of particles effectively trapped by the manifolds.
    This approach} was previously explored in \citet{Athanassoula2009b} in the context of a test-particle simulation of a simple analytical model. 
    Here, we build upon that initial approach and provide its first systematic and quantitative implementation in an $N$-body simulation.

    {This criterion allows us to define, at each time $t$, a set of trapped particles, $N_{\mathrm{trap}}(t)$.}
    The ratio of this trapped population to the total number of candidate particles provides a quantitative measure of the fraction of spiral-arm material whose dynamics are governed by the invariant manifolds.
    
    We adopt two complementary candidate selections, motivated by the results of Paper I.
    First, we compute the fraction of trapped particles relative to the manifold-compatible population, i.e. the subset of particles that can be directly influenced by the manifolds --we refer to this population as \textit{candidates}, $N_{\mathrm{cand}}(t)$:
    \begin{equation}
        f_{\mathrm{cand}}(t) = \frac{N_{\mathrm{trap}}(t)}{N_{\mathrm{cand}}(t)}\, .
    \end{equation}
    
    Second, we evaluate the trapped fraction relative to all particles that overlap with the configuration-space projection of the manifolds (and therefore with the spiral arms\footnote{{The overlapping between the configuration space projection of the invariant manifolds and the location of the spiral arms is not assumed a priori, but follows from the results of Paper~I (see Fig.~2b), on which the present analysis builds. This correspondence is further supported by the test performed in Sec.~\ref{sec:validation} of the present paper.}}), regardless of their Jacobi energy --we refer to this population as \textit{overdensity}, $N_{\mathrm{over}}(t)$:
    \begin{equation}
        f_{\mathrm{over}}(t) = \frac{N_{\mathrm{trap}}(t)}{N_{\mathrm{over}}(t)}\, .
    \end{equation}
    Taken together, these two estimators provide a more complete picture of the trapping process, allowing us to assess both the efficiency with which manifold-compatible particles are captured and the overall contribution of manifold trapping to the {entire} spiral-arm structure.

    \subsection{Validation of the criterion}\label{sec:validation}

    \begin{figure*}[!h]
       \centering
       \includegraphics[width=\hsize]{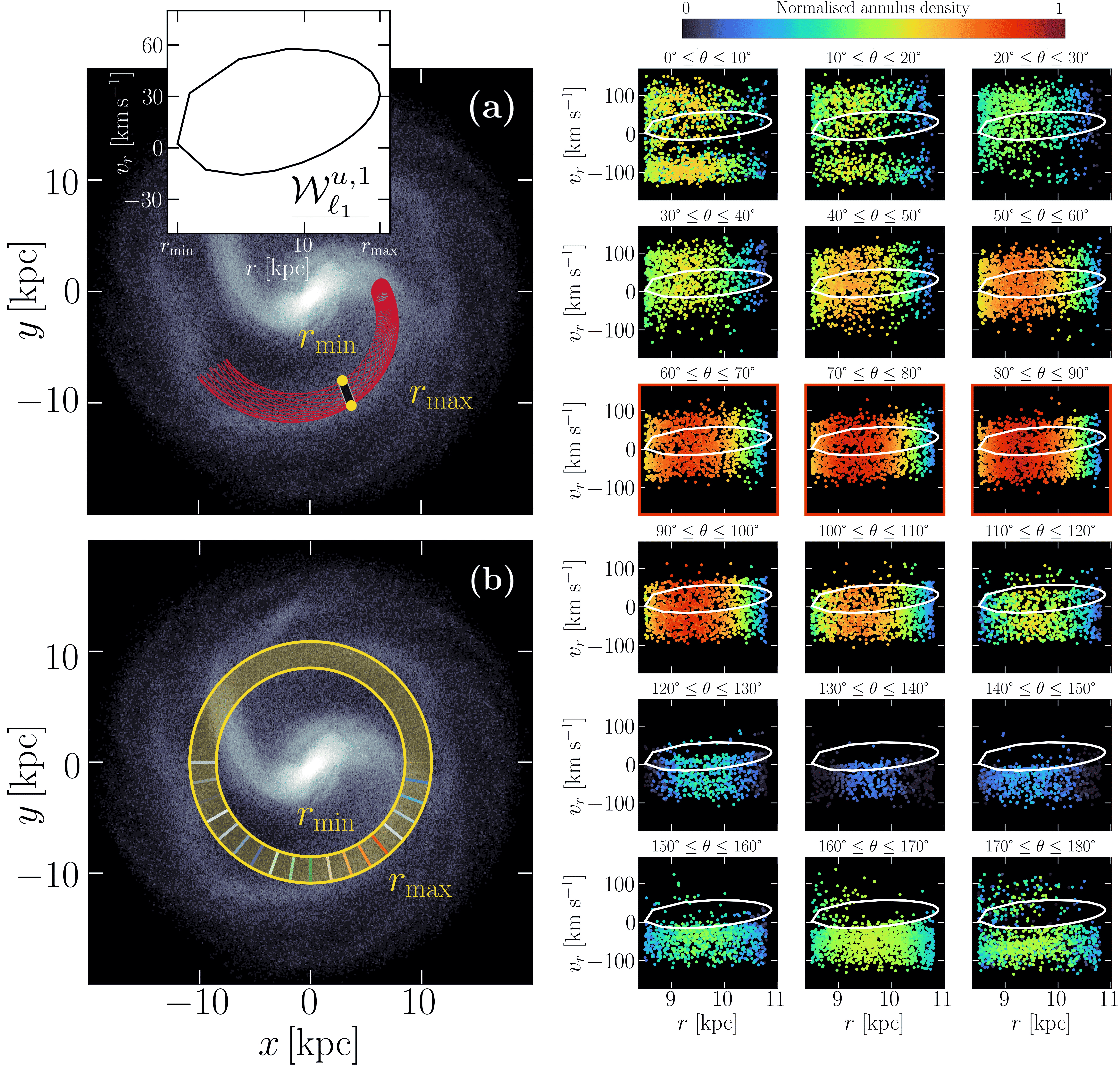}
            \caption{Analysis of the local effect of the exterior unstable branch of the invariant manifold associated with $\ell_1$ at its intersection with a section $\mathcal{S}$ at $\theta=70^\circ$, $\mathcal{W}_{\ell_1}^{u,1}$, after $t=0.853$ Gyr from the initial conditions.
            \textit{(a)}: Face-on view of the galaxy with the $\mathcal{W}_{\ell_1}^{u}$ manifold branch overplotted in red.
            Its intersection with the constant azimuth section $\mathcal{S}$ at $\theta=70^\circ$ is marked in black. 
            The yellow dots indicate the locations of the radial extrema of this intersection, $r_\text{min}$ and $r_\text{max}$.
            The inset panel shows this intersection, namely the $\mathcal{W}_{\ell_1}^{u,1}$ curve, in the $r-v_r$ plane.
            \textit{(b)}: Face-on view of the galaxy with a yellow annulus defined by $r_\text{min}$ and $r_\text{max}$.
            Azimuthal cuts on this annulus are indicated by coloured segments. 
            \textit{Right subpanels:} $r$–$v_r$ projection of all the manifold-compatible particles within the annulus in panel (b), binned by $\Delta\theta=10^\circ$ sectors. The white curve shows $\mathcal{W}_{\ell_1}^{u,1}$. The colour scale indicates the particle density within the annulus.
            Red boxes mark the angular sectors exhibiting the clearest intersection between the annulus and the spiral-arm overdensity.
            }
    
             \label{fig:B1_corba}
    \end{figure*}
    
    \begin{figure*}[!ht]
   \centering
   \includegraphics[width=\hsize]{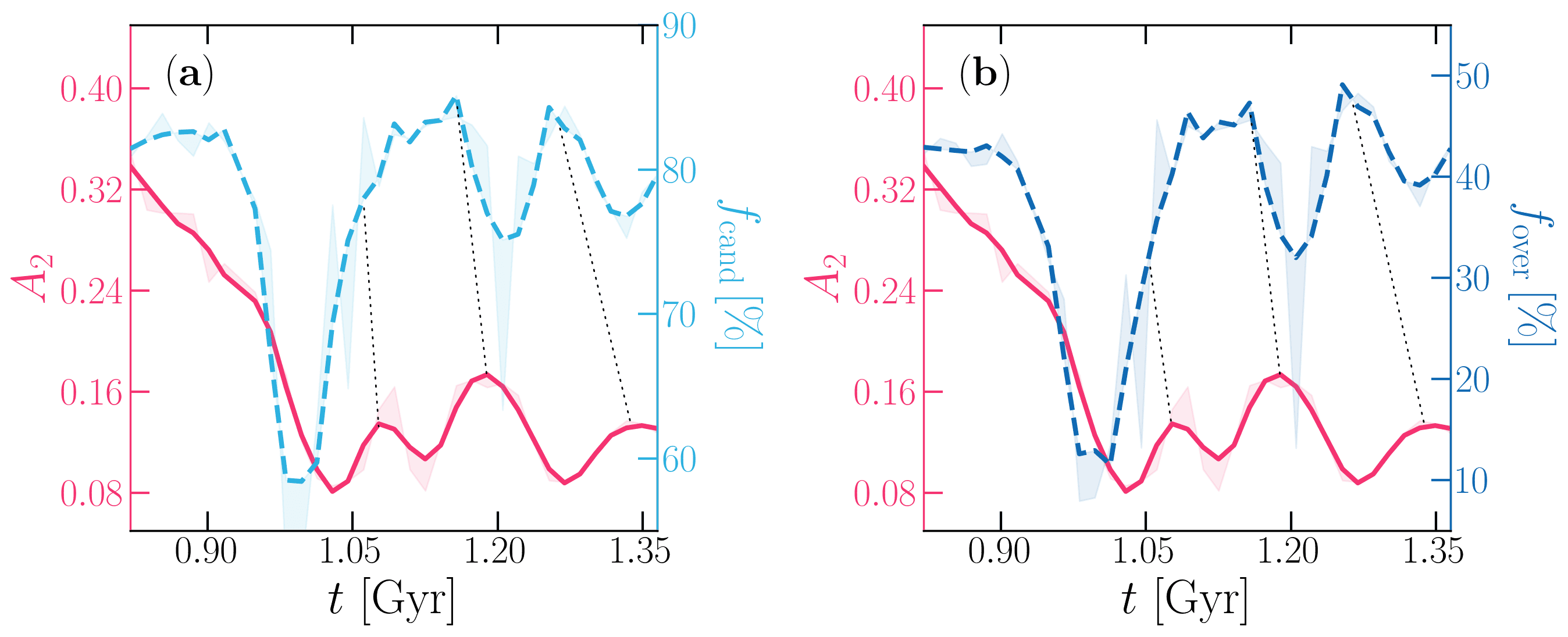}
        \caption{Time evolution of the fraction of trapped particles in the exterior unstable branches of the invariant manifolds (blue dashed curves) across $t=0.804$ Gyr and $t=1.365$ Gyr after the initial conditions, shown together with the spiral arm strength measured through the $A_2$ amplitude averaged over the inner spiral arm region, i.e. $[R_1,10]\,\text{kpc}$ --as shown in Fig. 1a in Paper I.
        \textit{Left}: the light-blue dashed curve shows the number of trapped particles over the number of (manifold-compatible) candidates.
        \textit{Right}: the navy-blue dashed curve shows the number of trapped particles over all the number of particles in the arms regardless of their Jacobi energy.
        Shaded regions correspond to deviations from the smoothed trends. 
        Black dotted lines highlight the correspondence between local maxima of the quantities shown in each panel.
        }

         \label{fig:B1_combined}
   \end{figure*}
    
    In this section we test the criterion defined in Sect. \ref{sec:criterion} in order to verify that it reliably identifies particles associated with the spiral arms, rather than selecting particles indiscriminately across the disc. 
    Fig.~\ref{fig:B1_corba} illustrates this validation using the unstable exterior branch of the manifold associated with $\ell_1$ (by symmetry, the behaviour is identical for the branch associated with $\ell_2$).
    
    First, we determine the radial bounds $r_{\min}$ and $r_{\max}$ (Fig.~\ref{fig:B1_corba}a, yellow dots) from a constant-azimuth section $\mathcal{S}$, which in this case corresponds to $\theta = 70^\circ$.
    These bounds are then used to construct an annulus spanning the full azimuthal range {(Fig.~\ref{fig:B1_corba}b, yellow ring)}.
    The manifold-compatible particles within this annulus are subsequently divided into azimuthal sectors of width $\Delta\theta=10^\circ$ {(Fig.~\ref{fig:B1_corba}b, coloured straight lines)}, ordered clockwise to follow the orientation of the branch associated with $L_1$.
    For each sector we display the distribution of radial velocity as a function of radius in the panels on the right-hand side of Fig.~\ref{fig:B1_corba}.
    The particles {in these panels} are colour-coded according to their density relative to the {global} density of the annulus (warmer colours indicate overdense regions, while colder colours correspond to lower densities).
    In each of these panels we also overplot the curve  $\mathcal{W}_{\ell_1}^{u,1}$ {(white curve)}, corresponding to the intersection of the manifold with the constant-azimuth section $\mathcal{S}$ at $\theta=70^\circ$ (shown in the inset panel of Fig.~\ref{fig:B1_corba}a).\\
    
    If the trapping criterion were not physically meaningful, the $\mathcal{W}_{\ell_1}^{u,1}$ curve would reproduce the distribution of particles in the $r-v_r$ plane irrespectively of azimuth.
    However, this behaviour is not observed.
    Instead, the curve provides a good match to the particle distribution only in sectors close to $\theta=70^\circ$ (Fig.~\ref{fig:B1_corba}, red-framed panels).
    In these panels the radial velocities cluster around the manifold curve $\mathcal{W}_{\ell_1}^{u,e}$ and coincide with the highest-density regions of the annulus, corresponding to the location of the spiral arm.
    As expected, a good agreement is not restricted to the bins at $60^\circ\leq \theta \leq70^\circ$ and $70^\circ\leq \theta \leq80^\circ$, but also extends to neighbouring azimuthal sectors. 
    This behaviour follows naturally from the continuity of both the spiral-arm structure (which overlaps with the defined annulus, see Fig. \ref{fig:B1_corba}b) and also from the continuity in radial velocity.
    At azimuths progressively farther from this region, the particle distribution departs from the manifold curve and the overdensity disappears, thus representing an inter-arm region. \\
    
    This behaviour demonstrates that the manifold-derived constraint is not generally satisfied by particles throughout the disc, but specifically characterises the phase-space structure associated with the spiral arms. 
    The azimuthal variation of the radial velocity distribution is also consistent with the velocity maps presented in Fig. 4h of Paper I.
    This test therefore confirms that the proposed criterion provides a robust and physically consistent phase-space diagnostic of manifold-trapping in the spiral arms. 
    
\section{Manifold-trapping in spiral arms}\label{sec:results}   
    \subsection{Spiral arm strength and trapped fraction}\label{sec:trapped_arms}
   We quantify the fraction of particles trapped within the unstable exterior branches of the invariant manifolds at each simulation snapshot, following the procedure described in Section~\ref{sec:criterion}. 
   The temporal evolution of these trapped fractions is shown in Fig.~\ref{fig:B1_combined}a (light-blue dashed curve: fraction with respect to candidates) and  Fig.~\ref{fig:B1_combined}b (navy-blue dashed curve: fraction with respect to all the kinematic populations in the arms, i.e. irrespective of their Jacobi energy).
   Both panels also compare these trapped fractions with the spiral arm strength traced by the $m=2$ Fourier amplitude $A_2$ (pink solid curve), radially averaged over $R_1-10$~kpc, where $R_1$ denotes the outer edge of the bar as defined in Appendix B of \citet{Dehnen2022}.

   Remarkably, neither the trapped fractions nor the spiral arm strength in Fig. \ref{fig:B1_combined} remain constant over time.
   In particular, the trapped fractions exhibit significant temporal variability, oscillating between $\sim 10-50\%$ taking into account all the kinematic populations in the arms, and between $\sim 60- 90\%$ for the candidate sample (i.e. from manifold-compatible particles).
   This supports the picture of spiral arms as transient, recurrent features rather than steady, long-lived structures, as reported in \citet{SellwoodCarlberg1984, Athanassoula2012, Baba2013} among others.
   Both panels in Fig.~\ref{fig:B1_combined} display very similar temporal trends.
   The systematic offset between the trapping fraction values shown in panels (a) and (b) simply reflects the relative abundances of the different kinematic populations in the disc, consistent with the results presented in Paper~I (see Fig.~5 therein).
   
   It should be noted that the percentage of manifold-trapped particles in the analysed timespan never fully vanishes.
   Instead there is a remaining non-zero baseline level ($\sim 10\%$ in the total overdensity, and $\sim 60\%$ of the manifold-compatible population) throughout the evolution, indicating that invariant manifolds provide a persistent dynamical framework that can sustain spiral structure, even during low-amplitude phases, as long as the bar is present and dominant (as shown in Fig. 1a of Paper I).

   A clear co-evolution is visible between the quantities in Fig.~\ref{fig:B1_combined}: maxima and minima of the trapped fractions (blue dashed curves) broadly coincide with variations in $A_2$ (pink solid curve). Black dotted lines further highlight the correspondence between local maxima in each panel.
   To quantify this relation, we computed the normalised cross-correlation function (CCF) between the smoothed, detrended time series. By construction, the CCF ranges from $-1$ (perfect anti-correlation) to $+1$ (perfect correlation), while values near $0$ indicate no correlation, as expected for unrelated or random signals.
   We find a maximum correlation of $\mathrm{CCF}_{\max}=0.7$ for both definitions of the trapped fraction (i.e. relative to the candidate sample and to all particles in the arms). This value is significantly higher than the level expected from purely stochastic fluctuations, which would produce correlations close to zero with no well-defined peak. 
   The presence of a strong, well-localised maximum in the CCF therefore indicates a statistically meaningful temporal coupling between the two quantities.
   In both cases, the CCF peaks at a lag of $\Delta t \approx 33\,\mathrm{Myr}$, with the trapped fraction leading and the spiral arm strength responding subsequently. 
   This lag corresponds to approximately two simulation snapshots, indicating a tight temporal connection between the two quantities. 
   The observed lead of the trapped fraction over $A_2$, although limited to a short delay, is physically consistent with {the material density wave} scenario {proposed in Paper I}. 
   {In this picture, manifold-compatible particles constitute the dynamical backbone of the spiral structure, while the low-energy population provides a surrounding density enhancement. 
   The resulting spiral pattern thus has a hybrid character: a genuinely dynamical component linked to manifold-driven particle transport, together with a density-wave-like response of the surrounding disc. 
   Within this framework, a temporal offset between these components is naturally expected, since the gravitational response of the disc is not instantaneous. 
   As particles are progressively trapped and guided along the manifolds, their collective self-gravity builds up over time, inducing a delayed response in the surrounding low-energy population. 
   The temporal offset identified here is therefore consistent with this causal sequence, connecting the dynamical trapping process to the subsequent amplification of the spiral pattern.
   See Appendix \ref{sec:appendix} for an explicit analysis of the self-gravity of the trapped population and its effect on the surrounding disc.}
   {This behaviour contrasts with the expectations from the classical density-wave scenario, where the spiral amplitude arises from an essentially instantaneous orbital response of the disc, and no clear temporal ordering is expected between the density enhancement and any specific subset of particles. 
   In our analysis, by construction, the trapped fraction includes only particles that are energetically compatible with the manifolds, whereas the Fourier amplitude $A_2$ includes contributions from all disc particles, regardless of their Jacobi energy. 
   The fact that the trapped fraction systematically precedes the variations in $A_2$ therefore points to a causal sequence in which particles are first captured into manifold-supported orbits, and the global spiral amplitude responds afterwards. 
   This small but systematic delay can be interpreted as the response time of the disc to the additional self-gravity of the trapped population (see Figs.~\ref{fig:self-gravity_1r_pic} and \ref{fig:self-gravity_4t_pic}).
   In this view, manifold trapping does not only confine particles moving along the arms, but also actively drives the growth of the spiral overdensity, which subsequently affects the surrounding disc.}\\
   
   A peak-by-peak analysis of Fig.~\ref{fig:B1_combined} yields a characteristic oscillation period of $T_{A_2}\approx 112\pm10$~Myr for $A_2$ and $T_\%\approx100\pm20$~Myr for both trapped fractions, confirming that all quantities vary on comparable timescales.
   These results indicate that short-term fluctuations in the strength of spiral arms are tightly coupled to variations in the trapped fraction, with a very slight delay.
   Such recurrent timescales are consistent with previous studies reporting spiral-arm lifetimes of order of a few hundred million years in dynamically evolving barred discs \citep{Wada2011, Sellwood2011, Grand2012a, RocaFabrega2013, Baba2015, Quinn2026}.

   \subsection{Azimuthal profiles}\label{sec:profiles}
   \begin{figure}[!t]
   \centering
   \includegraphics[width=0.9\hsize]{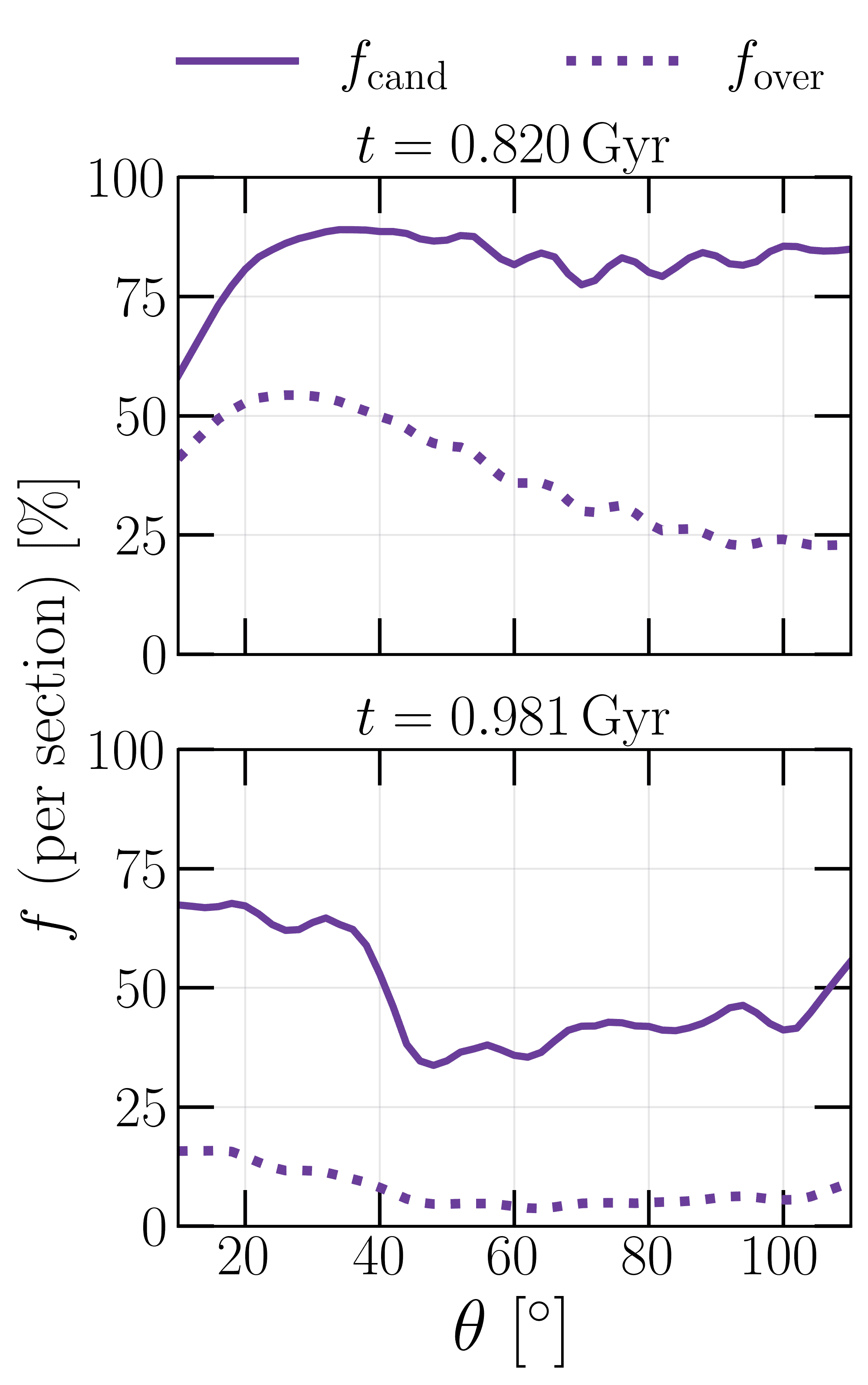}
        \caption{Azimuthal evolution of the trapped-particle fraction (per section) along the spiral arms for two selected snapshots.
        Both panels show the percentage of trapped particles by the unstable manifold branches at each section defined at $\theta$, measured along the spiral arms (both arms considered), with the solid line representing the trapped fraction with respect to the manifold-compatible population, and the dotted line relative to all the particles in the spiral arms regardless of their Jacobi energy.
        \textit{Top}: azimuthal profile at $t=0.820$ Gyr, maximum of the first trapping episode. 
        \textit{Bottom}: azimuthal profile at $t=0.981$ Gyr, minimum of the first trapping episode.}

         \label{fig:azim_evol}
   \end{figure}
   Complementary insight into the trapping process is provided by Fig. \ref{fig:azim_evol}, which presents the azimuthal distribution of trapped particles along the spiral arms for two representative snapshots. 
   Specifically, the azimuthal profiles correspond to the maximum and minimum trapped fractions associated with the first trapping episode identified in Fig. \ref{fig:B1_combined} ($t = 0.820$~Gyr, top panel; $t = 0.981$~Gyr, bottom panel). 
   In each panel, the solid curve shows the fraction of trapped particles relative to the manifold-compatible population, while the dotted curve represents the fraction relative to the total number of particles present in the spiral arms, irrespective of their Jacobi energy.
   
   At $t = 0.820$ Gyr, the azimuthal profile displays a broad, nearly plateau-like distribution, with trapping fractions remaining high across most of the analysed angular range. The trapped fraction relative to the manifold-compatible population stays close to $\sim80-90\%$, indicating that a large proportion of {manifold-compatible} particles are effectively confined by the unstable exterior branches throughout the arms' {extent}. 
   This behaviour is consistent with the strong spiral structure at this stage, when the $A_2$ amplitude reaches its highest values. 
   In contrast, the dotted curve shows a gradual decline with increasing azimuth, reflecting the underlying radial gradient in the density of manifold-compatible particles, which is higher in the inner arm regions and progressively decreases outward (see Paper I).
   Nonetheless, the effective trapping within the entire spiral arm overdensity (dotted curve, top panel) reaches values of up to $\sim55\%$ in certain azimuthal sections (up to $\theta=40^\circ$), implying that more than half of the particles in the spiral arms are manifold-trapped at these locations, thereby highlighting their significant dynamical contribution during phases of strong spiral structure.
   
   By $t = 0.981$ Gyr, the trapping efficiency is significantly reduced and the azimuthal profile becomes markedly more irregular. 
   The trapped fraction relative to the manifold-compatible population falls to $\sim35-70\%$ and exhibits a pronounced drop around $\theta \sim 40^\circ$. 
   The corresponding dotted curve remains low across the entire angular range, generally below $\sim15\%$, indicating that only a small fraction of all spiral-arm particles are trapped at this stage. 
   It is important to note, however, that the trapping fractions shown in Fig. \ref{fig:B1_combined}–\ref{fig:azim_evol} are normalised to the total number of particles (either from the candidate sample or the full overdensity), and therefore do not directly reflect the absolute decrease in particle numbers between the two snapshots. 
   The physical mechanism responsible for this sharp decline in trapping efficiency will be investigated in Paper III.
   
   Taken together, these profiles demonstrate that manifold trapping is not spatially uniform along the spiral arms. Rather, the trapping efficiency exhibits significant azimuthal variations and evolves over time in response to changes in the dynamical state of the bar–spiral system. 
   In particular, phases of strong spiral structure are associated with both higher and more spatially extended trapping fractions, whereas weaker spiral-arm configurations are associated with a decline in trapping efficiency and a more irregular azimuthal distribution. 
   This behaviour highlights the intrinsically time-dependent nature of manifold-driven dynamics, and suggests that the contribution of manifolds to spiral structure is modulated not only globally, but also locally along the arms.

   \subsection{Trapping persistence}\label{sec:IN--OUT}
   \begin{figure}[!b]
   \centering
   \includegraphics[width=\hsize]{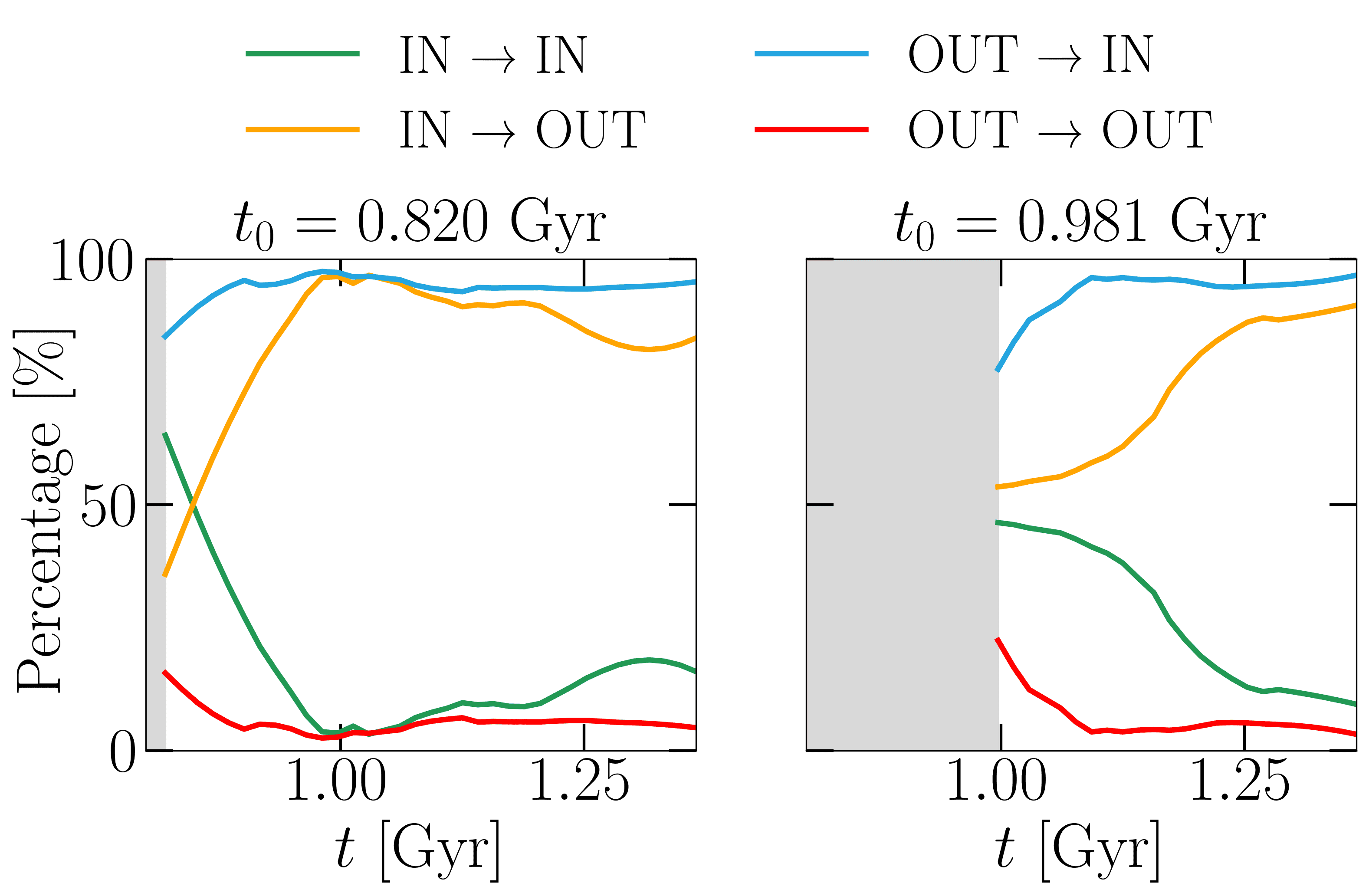}
        \caption{Temporal evolution of the dynamical state of particles initially classified as trapped (IN) or untrapped (OUT) by the unstable exterior branches of the invariant manifolds. 
        For two representative reference snapshots, we follow the subsequent evolution of the classification of the particles initially labelled at $t_0$. 
        \textit{Left}: $t_0 = 0.820$ Gyr, corresponding to the maximum of the first trapping episode.
        \textit{Right}: $t_0 = 0.981$ Gyr, corresponding to the minimum of the first trapping episode.
        In each panel, the curves show the fraction of particles that remain trapped (IN--IN, green), become untrapped (IN--OUT, yellow), become newly trapped (OUT--IN, blue), or remain unaffected by the manifolds (OUT--OUT, red).
        The shaded region indicates times earlier than the first snapshot for which measurements are available for each reference time.}

         \label{fig:IN--IN}
   \end{figure}

   \begin{figure*}[!t]
   \centering
   \includegraphics[width=\hsize]{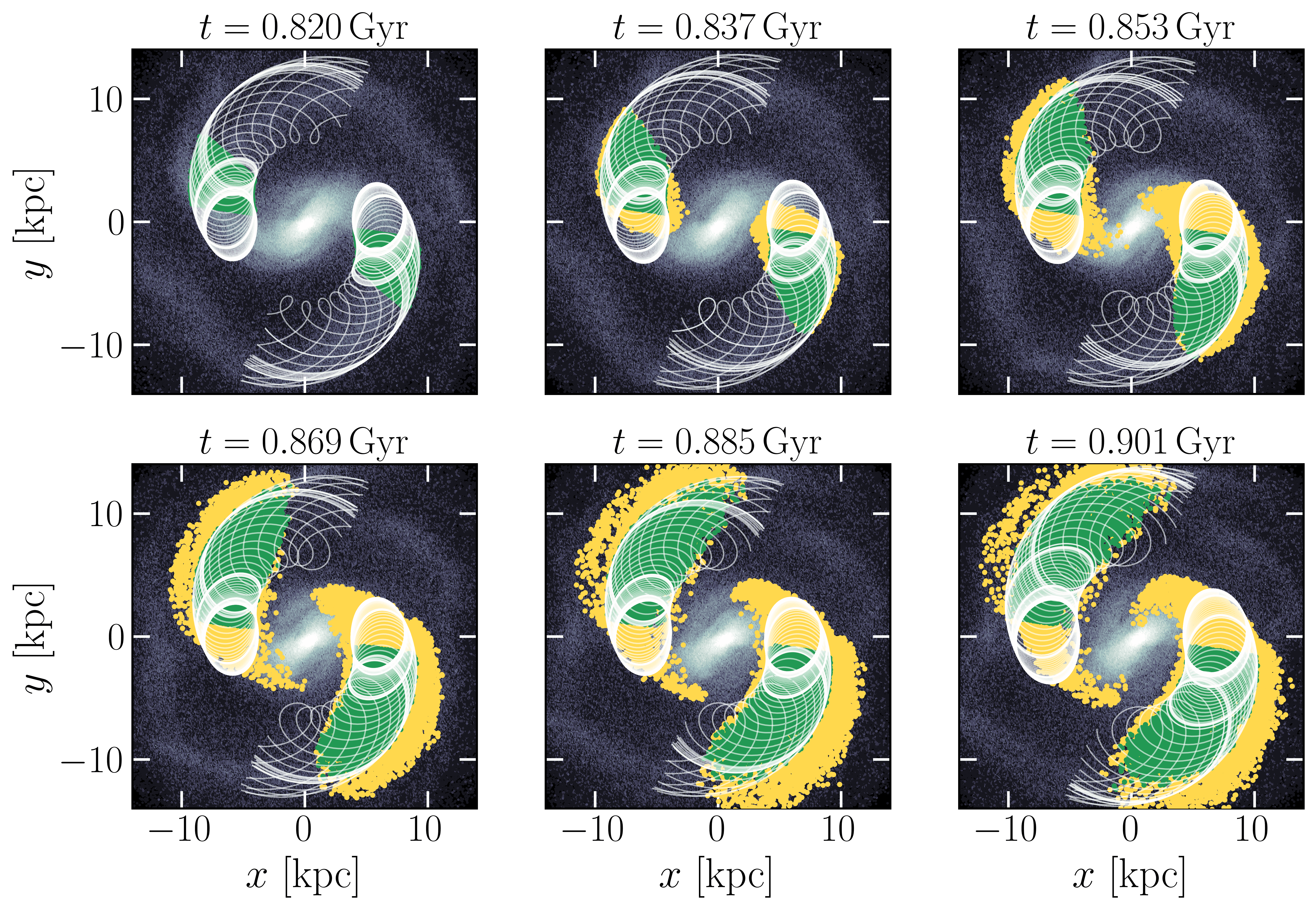}
        \caption{Time evolution of the bundle of trapped particles initially trapped at $t=0.820\,\mathrm{Gyr}$ within the angular interval $10^\circ \leq \theta \leq 40^\circ$.
        A total of 26,617 particles are overlaid in each panel (13,967 associated with the $L_1$ exterior unstable branch and 12,650 with the $L_2$ branch). The subsequent panels follow the same particles forward in time, illustrating the evolution of their trajectories and trapping state.
        Green dots correspond to particles that remain trapped (IN--IN) relative to the initial selection, while yellow dots indicate particles that escape from the trapped state (IN--OUT). White curves show the unstable branches of the invariant manifolds. The background displays the face-on surface density of the disc, and each panel is labelled with the corresponding simulation time.
        }

         \label{fig:track_in}
   \end{figure*}

   \begin{figure*}[!t]
   \centering
   \includegraphics[width=\hsize]{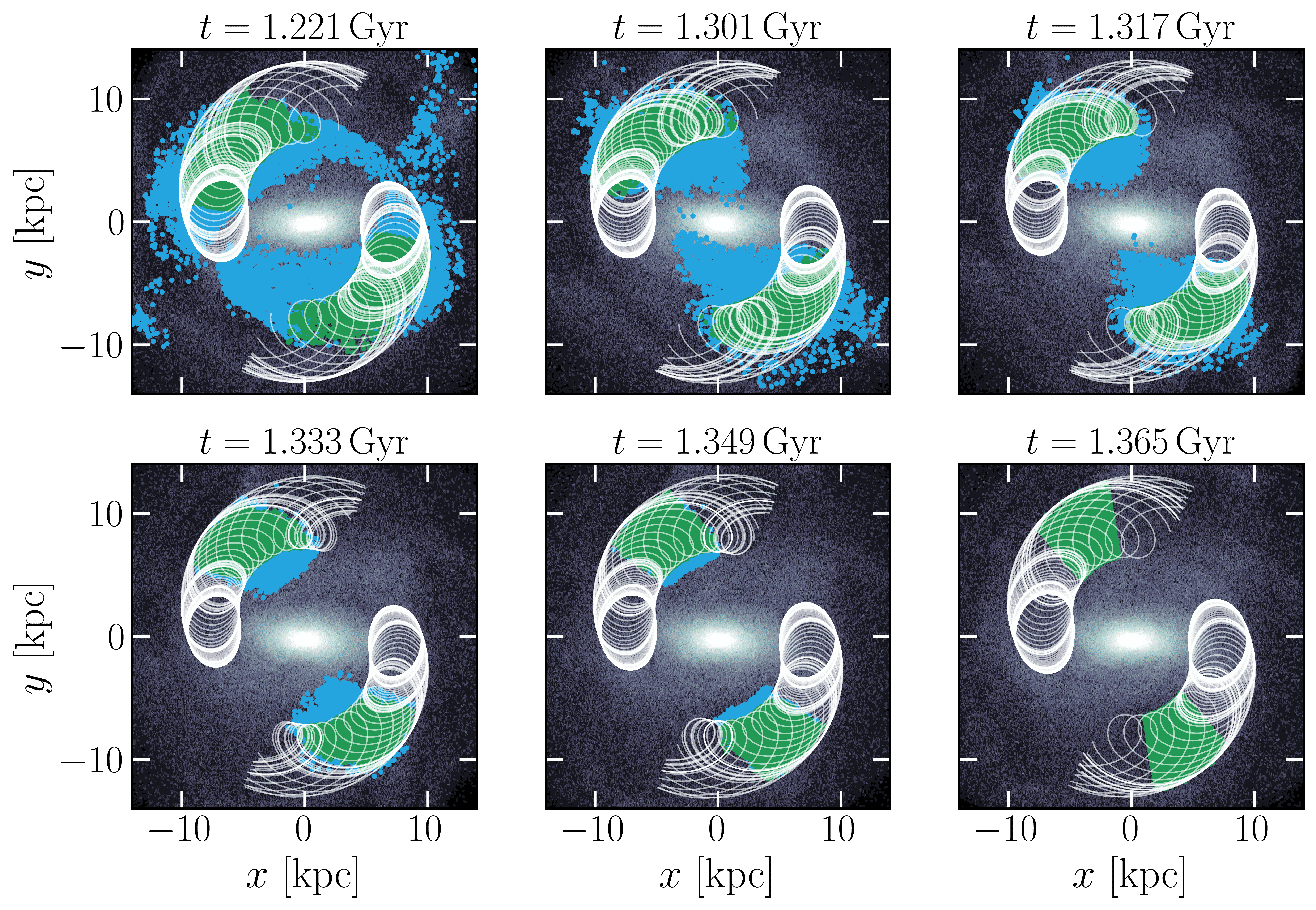}
        \caption{Time evolution of the bundle of trapped particles trapped at $t=1.365\,\mathrm{Gyr}$ within the angular interval $50^\circ \leq \theta \leq 80^\circ$ back in time.
        A total of 15,034 particles are overlaid in each panel (5,559 associated with the $L_1$ exterior unstable branch and 9,475 with the $L_2$ branch). The previous panels track the same particles backward in time, illustrating the evolution of their trajectories and trapping state.
        Green dots correspond to particles that were previously trapped as well (IN--IN), while blue dots correspond to particles that were initially untrapped but eventually follow the manifolds (OUT--IN). White curves show the unstable branches of the invariant manifolds. The background displays the face-on surface density of the disc, and each panel is labelled with the corresponding simulation time.
        }

         \label{fig:track_out}
   \end{figure*}

   Following this investigation, we analyse the dynamical evolution of particles initially classified as trapped (IN) or untrapped (OUT) at selected reference snapshots, in order to assess the temporal persistence of manifold trapping.
   Figure~\ref{fig:IN--IN} shows this analysis for the same two snapshots analysed in Fig.~\ref{fig:azim_evol} ($t_0=0.820$~Gyr, left panel; and $t_0=0.981$~Gyr, right panel), corresponding to the maximum and minimum of the first episode of manifold-trapping identified in Fig.~\ref{fig:B1_combined}.
   For each reference snapshot, we track the subsequent evolution of the same particle sample (i.e., those identified as trapped at $t_0$) and measure the fraction that remains trapped (IN--IN, green curve), becomes untrapped (IN--OUT, yellow curve), becomes newly trapped (OUT--IN, blue curve), or remains untrapped by the manifolds (OUT--OUT, red curve).
   Here, the classification into IN and OUT is defined according to our phase-space criterion (Sec. \ref{sec:criterion}): particles are considered trapped if, within the selected radial and azimuthal ranges (i.e. $r_{\min} \leq r \leq r_{\mathrm{man}}$, for each constant-azimuth section $\mathcal{S}$), their radial velocities lie within the closed curve $\mathcal{W}_{\ell_1}^{u,1}$ in the $(r, v_r)$ plane, with all quantities consistently evaluated at each snapshot, including the Jacobi energy. 
   Consequently, the OUT population does not account for all particles that are not trapped by the manifolds in the disc, but rather those that, within the same spatial constraints, do not satisfy the kinematic condition imposed by the manifolds.
   
   In both cases, Fig. \ref{fig:IN--IN} shows that the fraction of particles that remains trapped decreases with time.
   This behaviour is broadly consistent with the expected evolution of particles trapped by the unstable exterior branches of the invariant manifolds, which tend to propagate along the spiral arms from the innermost regions (i.e., near the saddle equilibrium points $L_1$–$L_2$) towards larger azimuths. 
   As a result, a fraction of the initially trapped population eventually leaves the analysed angular range, which extends up to the $\theta = 110^\circ$ section. 
   However, this effect alone does not fully account for the observed decline in the IN--IN. Additional factors—such as changes in the underlying dynamical structure, local perturbations, or deviations from the manifold-defined kinematic condition—may also cause particles to no longer satisfy the trapping criterion adopted in this work.
   A relevant quantity in this analysis is therefore the time required for the initially trapped population to cease being dominant.
   In Fig.~\ref{fig:IN--IN} this moment can be identified when the IN--IN fraction (green curve) drops to very low values.
   The characteristic trapping time inside the manifold flow is of the order of a few hundred Myr, which is comparable to the oscillation period of both the $A_2$ amplitude and the global trapped fraction (Fig.~\ref{fig:B1_combined}).
   The complementary behaviour is seen in the IN--OUT population (yellow lines, Fig.~\ref{fig:IN--IN}), whose fraction increases to $\sim90\%$, confirming that almost all initially trapped particles eventually escape from the manifolds as time progresses.
   
   This behaviour is illustrated in Fig.~\ref{fig:track_in}. Particles trapped at $t_0 = 0.820$ Gyr between the sections $\theta = 10^\circ$ and $\theta = 40^\circ$ are shown in the first panel. 
   This angular selection corresponds to the innermost angular sections where we quantify the trapping, as also shown in Fig. \ref{fig:azim_evol} (top panel). 
   Initially their positions are confined within the bounds of the unstable exterior manifold branches and the selected angular range (top left panel, Fig. \ref{fig:track_in}). 
   As time evolves, most particles continue to follow the manifolds towards larger azimuths (IN--IN, green dots), while an increasing fraction gradually transitions out of the trapped state (IN--OUT, yellow dots).
   As particles propagate along the spiral arms and spread along their full extent, the IN--IN population progressively becomes subdominant.
   It should be noted that in Fig.~\ref{fig:track_in} the green dots are plotted on top of the yellow ones for clarity.
   Importantly, the spatial distribution of IN--OUT particles still closely follows the projection of the invariant manifolds in configuration space, indicating that these particles remain dynamically influenced—i.e. guided—by the manifolds. 
   However, in our quantitative analysis we adopt a more restrictive definition of trapping based on phase-space criteria. 
   Under this conservative definition, proximity to the manifolds in configuration space alone is not sufficient to classify a particle as trapped, which explains why these guided particles are identified as IN--OUT in our framework.
   
   Interestingly, the IN--OUT particles exhibit two distinct behaviours. 
   Most of them escape into the disc while still roughly following the manifold contours, suggesting that their motion remains indirectly influenced by the manifold geometry, albeit in a less coherent manner. 
   A smaller fraction, however, migrates toward the bar region. This occurs because our trapping criterion is evaluated very close to the Lagrange saddle points ($\theta = 10^\circ$). In fact, these particles trace the unstable inner branches associated with the same equilibrium points (see Fig. \ref{fig:B1_sketch}).
   Nevertheless, the effect of these inner branches to the spiral-arm evolution lies beyond the scope of this work and will be examined in Paper III.
   In any case, Fig. \ref{fig:track_in} exemplifies the quantitative behaviour shown in Fig. \ref{fig:IN--IN} for particles initially trapped: some of them maintain their motion within the dynamical bounds of the manifolds, but their contribution gradually gets diluted as they move away from the region where they were initially trapped.\\

   At the same time, a large fraction of particles initially outside the manifolds subsequently becomes trapped\footnote{Note that the particles initially considered to be not trapped (OUT) are not all the disc particles that are not guided by the manifold, but those that, at each azimuth, their radial velocity does not lie within the $\mathcal{W}_{\ell_1}^{u,1}$ curve bounded radially by $r_\text{min}$ and $r_\text{max}$, as described in Sec. \ref{sec:methodology}.} (Fig. \ref{fig:IN--IN}, OUT--IN), reaching values up to $\sim90$–$100\%$.
   This demonstrates that the manifolds continuously capture new particles. Only a small fraction of particles remains outside the manifolds despite being dynamically compatible with them (OUT--OUT), typically $\lesssim10\%$, indicating that the invariant manifolds dominate the orbital dynamics in this region of phase space.
   Similarly to Fig.~\ref{fig:track_in}, Fig.~\ref{fig:track_out} illustrates the origin of particles at $t=1.365$ Gyr that are trapped between sections $\theta=50^\circ$ and $\theta=80^\circ$.
   In this case, the selection of trapped particles (lower right panel, Fig.~\ref{fig:track_out}) are tracked backwards in time.
   As we move backwards, the particles that were also previously trapped are distributed along lower azimuths.
   The most interesting aspect of this figure is the behaviour of the particles that were previously not trapped but are later on trapped at $t=1.365$ Gyr.
   Some of these particles come from the disc, which can be seen at radii larger than the radial extent of the manifolds.
   Nonetheless, most of them come from regions close to the bar, particularly from the stable interior branches associated with each saddle point (see Fig. \ref{fig:B1_sketch}).
   Although the trapping or indirect influence of these other branches of the invariant manifolds is not analysed in this work, the trajectories of these particles reveal that these branches also contribute to the overall picture of spiral arms formation.

   All in all, Figs.~\ref{fig:track_in} and \ref{fig:track_out} show that, although there exists a stream of particles trapped within the invariant-manifold dynamics that constitutes the backbone of the spiral arms, there is also a more complex interplay with manifold-compatible particles, which can dynamically become trapped and untrapped. 
   Both examples illustrate the influence that the trapped population exerts on its surroundings, affecting particles with compatible energies through both the manifold dynamics and self-gravity. 
   This process leads to their capture within the manifolds and contributes to the subsequent strengthening of the spiral structure. 
   This interpretation is also consistent with the material density wave described in Paper I.
   
   Overall, the behaviour shown in Figs.~\ref{fig:IN--IN}–\ref{fig:track_out} highlights the dynamical exchange between the trapped and untrapped populations and reinforces the picture suggested by Fig.~\ref{fig:B1_combined}: manifold trapping is an intrinsically time-dependent process. 
   These results demonstrate that spiral arms supported by invariant manifolds are not populated by a fixed set of particles. 
   
   Instead, these spiral arms are sustained by a combination of particles that remain temporarily trapped, particles that are progressively injected into and escape from the manifold phase-space tubes, and others that follow the dynamical flow of the manifolds without necessarily satisfying the adopted trapping criterion at all times. 
   In this sense, the spiral structure corresponds to a continuous and evolving phase-space circulation. 
   The observed oscillations in the trapped fraction therefore arise naturally from variations in the rate at which particles are injected into, and escape from, the manifold-supported structures.

   \subsection{Caveats and limitations}\label{sec:caveats}
   
   The analysis presented in this work for the quantification of the impact of invariant manifolds on the formation of spiral arms is subject to two main limitations.
   
   First, the criterion adopted in this work focuses exclusively on the role of the exterior unstable branches of the manifolds.
   While this choice is physically motivated, it provides only a partial description of the manifold-driven dynamics. 
   As illustrated in Figs. \ref{fig:track_in}–\ref{fig:track_out}, the remaining manifold branches also contribute to the overall structure, albeit at a secondary level.
   A more complete characterisation would therefore require incorporating all branches in order to fully capture the global manifold influence.
   
   Second, the notion of trapped particles, central in our criterion, is considerably more subtle in $N$-body simulations than in the test-particle framework on which our approach is based \citep{MRG2006, Athanassoula2009b}. 
   In self-consistent systems such as the presented in this study, particle self-gravity introduces additional complexity that slightly hinders the distinction between trapped and non-trapped populations. 
   Indeed, our analysis reveals the presence of particles that do not formally satisfy the trapping criterion—such as those in the IN–OUT category—yet remain closely aligned with the manifolds in configuration space and continue to follow their large-scale flow (see Fig. \ref{fig:track_in}). 
   This behaviour suggests that invariant manifolds exert a broader dynamical influence, extending beyond strictly trapped particles through indirect effects, such as the self-gravity of the overdensities they support, consistent with the description of \citet{Gomez2004} (see there Fig. 6), where the influence of invariant manifolds is shown to extend beyond transit orbits, also affecting non-transit orbits at a more local level.

   In this sense, the present criterion should be regarded as a first step toward quantifying the role of invariant manifolds in fully self-consistent systems. 
   A more complete understanding will require moving beyond this first-order description and explicitly accounting for second-order, indirect effects induced by the manifolds on particles that are not formally trapped. 
   Such an approach would provide an even more global and physically comprehensive picture of their overall dynamical impact.

\section{Conclusions}\label{sec:conclusions}
In this work, we have quantified, for the first time, the effect of exterior unstable branches of invariant manifolds in the formation of spiral arms in a fully self-consistent $N$-body simulation of a barred galaxy.
We have defined a criterion to quantify the direct influence that these invariant manifolds branches exert on particles in the spiral arms.
In a nutshell, a particle is considered to be \textit{trapped} by them if it simultaneously satisfies three conditions: \textit{(i)} its Jacobi energy lies within the manifold-compatible range --i.e. $E_{L_{1,2}}\leq E_J\leq E_\text{man}$, \textit{(ii)} its radial position falls within the {bounds set by the projection of the manifolds onto configuration space} at a given azimuth, and \textit{(iii)} its radial velocity lies inside the corresponding $r$–$v_r$ curve.
After applying this criterion to the snapshots in this simulation where two bisymmetric, grand-design spiral arms are distinguishable, we derive the following main conclusions:
   \begin{enumerate}
      \item Invariant manifolds play a significant role in the formation and maintenance of spiral arms in barred galaxies. In this simulation, we find that between $10-50\%$ of all the particles located in the spiral-arm region are guided by the unstable exterior branches of the manifolds.
      This fraction raises up to $60-90\%$ when considering manifold-compatible particles, that is, those whose Jacobi energy is bounded by the energy of the equilibrium points $L_1-L_2$ and the energy of the manifolds.
      These results are broadly consistent with the early estimates reported by \citet{Athanassoula2009b}. 
      While those estimates were inferred from a statistical classification of orbits in analytic barred--spiral potentials, our measurements are based on a systematic census of particles across successive snapshots of a fully self-consistent $N$-body simulation, capturing the time-dependent gravitational field and the non-linear response of the disc. 
      The remaining spiral-arm particles are not manifold-trapped in a strict sense; however, in our framework, manifold-driven flows can naturally induce density-wave-like behaviour through the self-gravity of the manifold-guided overdensity, which we referred to as material density waves in Paper I.\\
      
      \item The strength of the spiral structure is tightly coupled to the efficiency of manifold trapping: periods of enhanced spiral Fourier $m=2$ amplitude {correlate} with phases in which a larger fraction of particles are confined within the unstable exterior branches of the manifolds.
      Both quantities rise and decay {with a small but systematic temporal delay, of order $\sim 50$ Myr in our model, with variations in the trapped fraction preceding the response in spiral strength, $A_2$. 
      Within our framework, this delay can be naturally interpreted in terms of the material density waves scenario proposed in Paper I.
      Invariant manifolds first trap manifold-compatible particles, which then stream along the spiral arms. The self-gravity of this material subsequently amplifies the associated overdensities and induces a collective response in particles that are not necessarily manifold-compatible, leading to the observed increase in $A_2$.}\\

      \item We find that variations in the fraction of manifold-trapped particles are systematically associated with the growth of the spiral arms, with peaks in trapping preceding or coinciding with enhancements in spiral strength. These fluctuations occur on characteristic timescales of $\sim 100$~Myr, in good agreement with reported spiral-arm variability in dynamically evolving discs \citep{Wada2011, Sellwood2011, Grand2012a, Grand2012b, RocaFabrega2013, Baba2015, Quinn2026}. 
      {Beyond its temporal variability, the trapping efficiency also exhibits significant azimuthal variations along the spiral arms, reflecting the inherently non-axisymmetric and evolving nature of the manifold structure.}
      Together, these results support a picture in which manifold-driven particle trapping plays an active dynamical role in triggering or reinforcing transient, recurrent spiral structure.\\

      \item Importantly, the trapped fraction never vanishes, remaining at a non-zero baseline level ($\sim10\%$ in the total overdensity, and $\sim60\%$ of the manifold-compatible population) throughout the evolution, indicating that invariant manifolds provide a persistent dynamical backbone that sustains spiral structure, even during low-amplitude phases, as long as the bar is present.\\
   \end{enumerate}
In light of the above, we conclude that invariant manifolds play a significant role in shaping and sustaining spiral arms in barred galaxies. 
While our results are based on a single $N$-body simulation, they provide a clear proof of concept and a solid starting point for future work.
{However, the direct applicability of the results in this work is formally restricted to systems sharing similar dynamical properties.}
In forthcoming studies, we will extend this analysis to a broader suite of simulations spanning different initial conditions and numerical resolutions, and will explore more realistic physical prescriptions, including hydrodynamical models with gas, stellar ages, and chemical abundances. 
In the next paper of this series (Paper III), we will further investigate the physical origin of the observed variability in manifold trapping and its connection to the transient nature of spiral arms. In particular, we will aim to identify the mechanisms that regulate the strengthening and weakening of the manifold-supported structures, providing a dynamical interpretation of why prominent spiral arms can naturally fade and re-emerge over time within the invariant manifolds framework.

\begin{acknowledgements}
      T.S.T. acknowledges that this work was partially supported by the FI-STEP predoctoral grant program of the Department of Research and Universities of the Generalitat de Catalunya, co-funded by the European Social Fund Plus (reference number: 2025 STEP 00196). 
      M.R.G. and T.S.T acknowledge that this work was (partially) supported by the Spanish MICIN/AEI/10.13039/501100011033 and by "ERDF A way of making Europe" by the European Union through grants PID2021-122842OB-C21 and PID2024-157964OB-C21, the Institute of Cosmos Sciences University of Barcelona (ICCUB, Unidad de Excelencia María de Maeztu) through grant CEX2024-001451-M and the project 2021-SGR-00679 GRC de l’Agència de Gestió d’Ajuts Universitaris i de Recerca (Generalitat de Catalunya). 
      S.R.F. acknowledges the finantial support by the Swedish National Space Agency Senior career grant with number 2025-00181, and the Spanish Ministry of Science and Innovation through the research grants: PID2021-123417OBI00, funded by MCIN/AEI/10.13039/501100011033/FEDER, EU; PCI2022-135023-2, funded by MCIN/AEI/10.13039/501100011033 and the EU “NextGenerationEU” / PRTR; and PID2024-157374OBI00, funded by MI-CIU/AEI/10.13039/501100011033/FEDER, EU.
\end{acknowledgements}

\bibliographystyle{aa}
\bibliography{bibliography}

\begin{appendix}
\section{Gravitational impact of manifold-trapped material on the disc}\label{sec:appendix}

    The temporal offset between the maxima of the trapped fraction and the spiral Fourier amplitude $A_2$ reported in Fig.~\ref{fig:B1_combined} can be understood as a consequence of the self-gravity generated by the manifold-trapped population. To investigate this connection, we computed the gravitational potential directly from the full $N$-body particle distribution at different snapshots, without imposing any symmetry, smoothing prescription, or Fourier decomposition. 
    The purpose of this analysis is to determine whether the material channelled through the exterior unstable manifold branches, $\mathcal{W}_{\ell_i}^u$, with $i\in\{1,2\}$, generates a non-axisymmetric gravitational perturbation capable of driving the subsequent amplification of the spiral structure.\\

\begin{figure*}[!h]
       \centering
       \includegraphics[width=0.91\hsize]{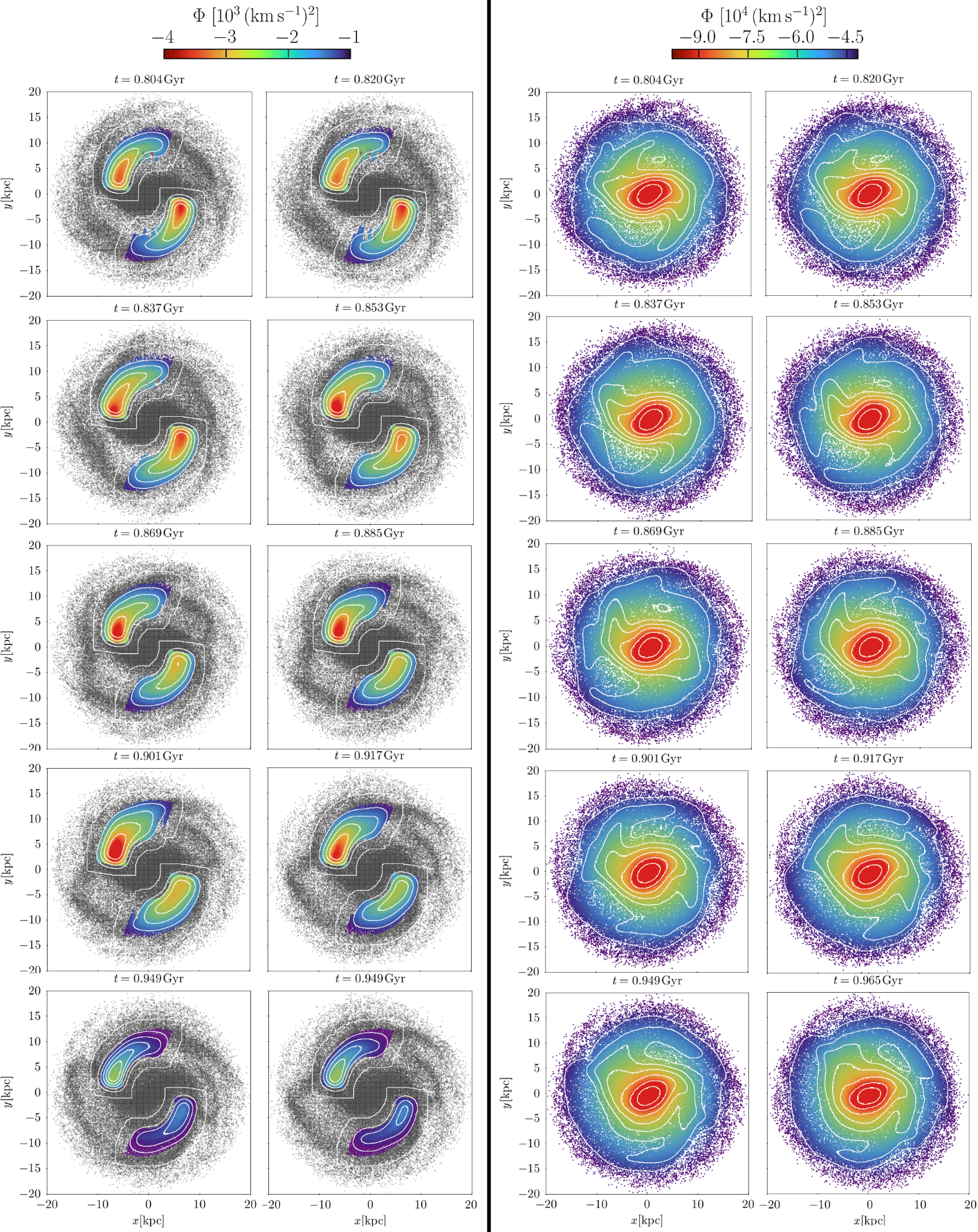}
            \caption{
            Time evolution of the gravitational potential contribution of the manifold-trapped population (left panels) and the contribution of the remaining particle population, i.e. the potential generated by all particles not classified as manifold-trapped (right panels), computed self-consistently using a brute-force method. Colours indicate the value of $\Phi$, with warmer tones corresponding to deeper potential wells and colder tones to higher potential values. White curves show equipotential contours.
            In the left panels, the grey background dots shows the full stellar particle distribution of the galaxy at each time for reference.
            Each panel presents a face-on view of the disc for ten consecutive snapshots spanning the interval between $t=0.804$~Gyr and $t=0.965$~Gyr.
            This sequence corresponds to the first trapping episode identified in Fig.~\ref{fig:B1_combined}.}
             \label{fig:self-gravity_1r_pic}
    \end{figure*}

    For each snapshot, we identified the particles trapped according to the criterion described in Sec.~\ref{sec:criterion} and computed their gravitational contribution self-consistently through a brute-force evaluation of the potential, subsequently averaged on a regular mesh with $100\,\mathrm{pc}\times100\,\mathrm{pc}$ cells spanning a $20\,\mathrm{kpc}\times20\,\mathrm{kpc}$ grid.
    The left panels in Figures~\ref{fig:self-gravity_1r_pic} and \ref{fig:self-gravity_4t_pic} show the temporal evolution of this potential during the first and last trapping episodes identified in Fig.~\ref{fig:B1_combined}. 
    The manifold-trapped population contains of the order of $\sim6\times10^4$ particles, out of a total stellar component of $10^6$ particles.
    On the other hand, the right panels in these figures show the potential generated by the complementary particle distribution, namely all particles not classified as trapped according to our criterion, including the dark matter halo component -- which makes the resulting potential systematically deeper than that shown in the left panels.
    A random subset of $10^5$ particles was selected to evaluate the potential associated with this much larger component.

    In Fig.~\ref{fig:self-gravity_1r_pic}, which corresponds to the first trapping episode in Fig.~\ref{fig:B1_combined}, the spiral pattern is already strongly developed at the beginning of the sequence and weakens with time. 
    This evolution is lead by the decrease in the trapped fraction, indicating that the efficiency of manifold-trapping declines during the episode.
    As a consequence, the non-axisymmetric gravitational perturbation generated by the trapped population also becomes weaker. 
    This trend is clearly visible in the left panels of Fig.~\ref{fig:self-gravity_1r_pic}, where the deep potential wells initially associated with the manifold branches gradually become shallower and less extended as the trapped fraction decreases.
    The weakening of the spiral arms therefore appears closely correlated with the reduction of the self-gravitating contribution produced by the trapped material. 
    In other words, as fewer particles remain trapped along the unstable exterior manifold branches, the corresponding overdensities become less prominent, reducing the amplitude of the associated spiral perturbation. 
    The physical origin of the weakening of the spiral structure during this first episode will be analysed in detail in Paper~III of this series. 
    However, the behaviour shown in Fig.~\ref{fig:self-gravity_1r_pic} already provides clear evidence that the effectiveness of manifold trapping has a direct impact on the depth of the non-axisymmetric potential wells and, consequently, on the strength of the spiral arms.
    
    Fig.~\ref{fig:self-gravity_4t_pic} presents this analysis for the last trapping episode identified in Fig.~\ref{fig:B1_combined}.
    The main difference with respect to Fig.~\ref{fig:self-gravity_1r_pic} --where the spiral arms are very strong-- is that in this other regime the bar dominates over the spiral-arm contribution to the potential.
    As a consequence, the non-axisymmetric component associated with the spiral structure is comparatively weaker and spans a narrower dynamic range.
    For this reason, the values of the potential in the right panels of Fig.~\ref{fig:self-gravity_4t_pic} are displayed on a logarithmic scale, which enhances the visibility of these low-amplitude perturbations and makes their spatial structure easier to identify.
    The left panels of Fig.~\ref{fig:self-gravity_4t_pic} also reveal a clear and systematic evolution of the non-axisymmetric gravitational perturbation produced by the trapped material. 
    As particles accumulate along the exterior unstable manifold branches, progressively deeper and more coherent potential wells develop along the spiral arms. This behaviour is particularly evident in the blue-framed panel, which corresponds to the trapped fraction maximum during this trapping episode ($t=1.269$~Gyr). 
    At this time, the low-potential regions associated with the efficiency of manifold-trapping reach their maximum spatial extent and depth, indicating that the accumulation of trapped material generates a significant non-axisymmetric self-gravitating perturbation. 
    Importantly, this structure is not imposed by construction, but emerges naturally from the evolving particle distribution itself.
    
    The dynamical response of the rest of the disc can be followed in the right panels of Fig.~\ref{fig:self-gravity_4t_pic}. 
    Importantly, the onset of the large-scale spiral response occurs immediately after the trapped population reaches its maximum concentration at $t=1.269$~Gyr (blue-framed panel). At this stage, the overdensities generated by the manifold-trapped material already define the loci around which the spiral structure subsequently develops, while the surrounding non-trapped particles are still only weakly organised.
    In the subsequent snapshots, however, the non-trapped component progressively accumulates around these same regions, making the spiral perturbation increasingly coherent and extended across the disc.
    This delayed growth strongly suggests that the trapped population acts as the dynamical seed of the later global response: invariant manifolds first redistribute matter along preferred directions in phase space, creating localised overdensities and associated non-axisymmetric potential wells, and only afterwards does the rest of the stellar disc react collectively to this perturbation on a finite dynamical timescale.
    This evolution is particularly clear in the left spiral arm of Fig.~\ref{fig:self-gravity_4t_pic}. 
    In this episode, the gravitational well generated by the particles trapped in $\mathcal{W}_{\ell_2}^u$ is significantly deeper than that associated with $\mathcal{W}_{\ell_1}^u$, leading to a more pronounced non-axisymmetric perturbation on the left side of the disc. 
    Correspondingly, after $t=1.269$~Gyr, the left arm becomes progressively more extended and prominent, indicating that the surrounding non-trapped particles are being preferentially attracted and reorganised around the stronger self-gravitating perturbation generated by the trapped material.
    Therefore, the strongest global spiral response is reached later, in the magenta-framed panel corresponding to the maximum Fourier $A_2$ amplitude at $t=1.349$~Gyr, when the spiral structure has already developed across the disc.\\
    
    In conclusion, the temporal delay between the blue- and magenta-framed panels therefore reflects the time required for the bulk stellar population -- dominated by low-energy particles ($E_J<E_{L_{1,2}}$) -- to develop a coherent gravitational response to the perturbation generated by the manifold-trapped material.
    See Fig.~4g in Paper~I to further check the kinematic imprint of this population.
    Within this framework, the amplification of the global $m=2$ spiral mode is not instantaneous, but emerges as the delayed collective response of the disc to the self-gravity of the trapped population. The measured delay of approximately $\sim 50$~Myr between the maxima of the trapped fraction and those of $A_2$ is naturally consistent with the timescale over which the disc reorganises to amplify the spiral pattern.
    These results therefore support the interpretation that invariant manifolds constitute the underlying dynamical backbone of spiral structure in barred galaxies.

\subsection{Second-order effects}
    A second-order effect that may contribute to the delay between the peak in the trapped fraction and the maximum of the $A_2$ Fourier amplitude is the presence of small asymmetries between the populations trapped in $\mathcal{W}_{\ell_1}^u$ and $\mathcal{W}_{\ell_2}^u$. 
    For simplicity, the trapped fraction has been defined jointly for both manifold branches in order to compare directly with the global bisymmetric amplitude $A_2$. 
    In the ideal symmetric case, where both branches contain comparable trapped populations, the non-axisymmetric perturbation generated by the trapped material is expected to grow nearly synchronously in both spiral arms, and the resulting delay between the trapped fraction and $A_2$ maxima should therefore approach its minimum value.
    
\begin{figure*}[!h]
       \centering
       \includegraphics[width=0.91\hsize]{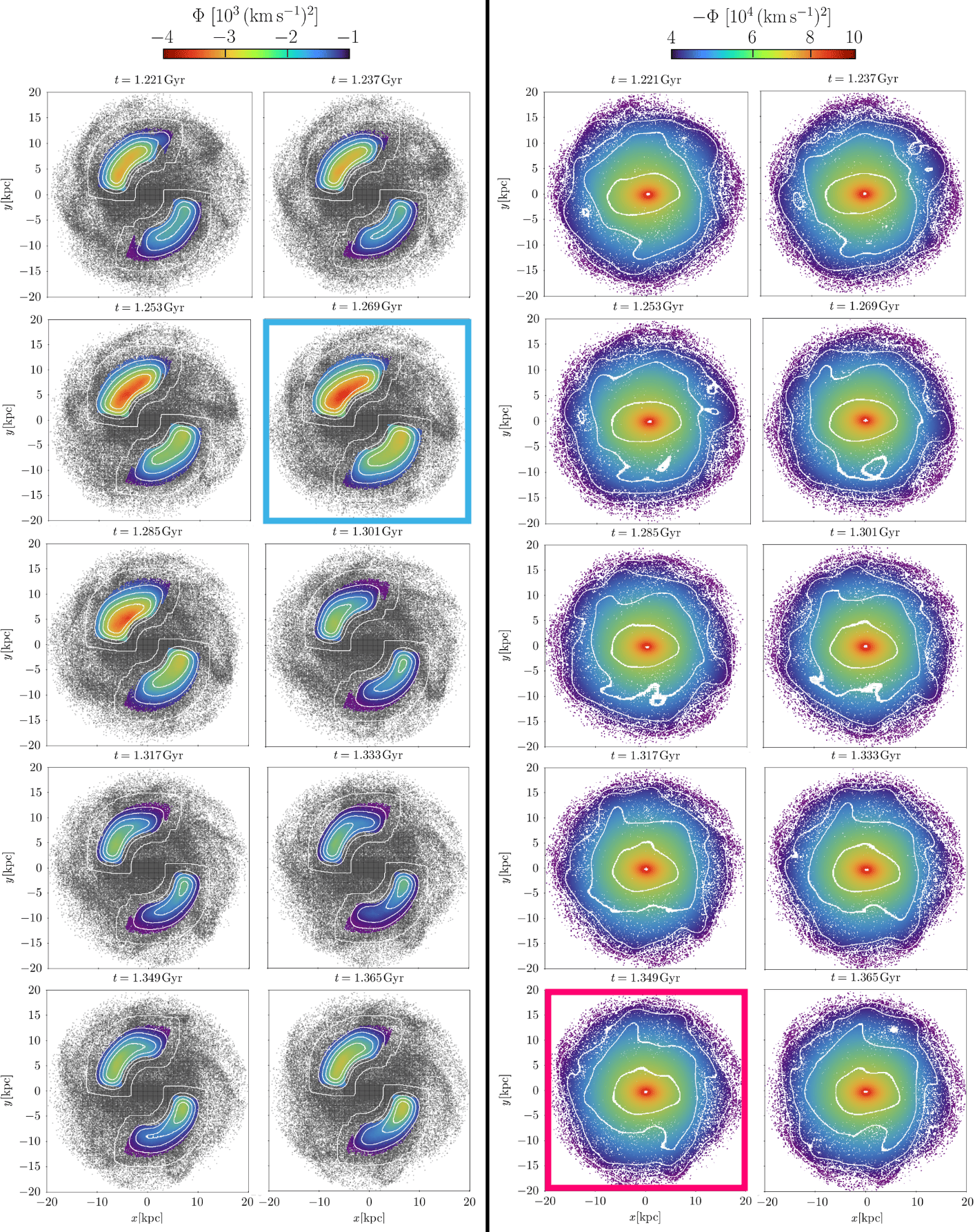}
            \caption{Time evolution of the gravitational potential contribution of the manifold-trapped population (left panels) and the contribution of the remaining particle population, i.e. the potential generated by all particles not classified as manifold-trapped (right panels), computed self-consistently using a brute-force method. Colours indicate the value of $\Phi$; for the right panels, the potential is displayed on a logarithmic scale. 
            Warmer tones correspond to deeper potential wells and colder tones to higher potential values. 
            White curves show equipotential contours.
            In the left panels, the grey background dots shows the full stellar particle distribution of the galaxy at each time for reference.
            Each panel presents a face-on view of the disc for ten consecutive snapshots spanning the interval between the maximum trapped fraction (blue-framed panel; $t=1.269$~Gyr) and the subsequent maximum of the spiral Fourier amplitude $A_2$ (magenta-framed panel; $t=1.349$~Gyr).
            This sequence corresponds to the last trapping episode identified in Fig.~\ref{fig:B1_combined}.}
    
             \label{fig:self-gravity_4t_pic}
    \end{figure*}

    However, if one manifold branch contains fewer trapped particles than the other, the associated self-gravitating potential well is comparatively weaker, causing the surrounding disc response to develop more slowly along that arm. 
    This behaviour is visible in Fig.~\ref{fig:self-gravity_4t_pic}, where the spiral arm associated with the weaker trapped population (right arm) develops slightly less rapidly than its counterpart. 
    Under these circumstances, the global $m=2$ mode only becomes fully coherent once both arms have sufficiently amplified, naturally introducing an additional contribution to the temporal offset between the trapped-fraction and $A_2$ maxima. 
    As a consequence, asymmetries between the trapped populations can maintain or slightly increase the observed delay.
    
    Importantly, however, such asymmetries cannot produce a situation in which the global $A_2$ maximum precedes the trapped-fraction peak. 
    The response of the low-energy stellar component is purely gravitational and therefore requires a finite dynamical timescale to develop after the non-axisymmetric perturbation generated by the trapped material has reached sufficient strength. 
    An asymmetry between the material trapped by the manifold branches $\mathcal{W}_{\ell_1}^u$ and $\mathcal{W}_{\ell_2}^u$ may delay the emergence of a coherent bisymmetric response, but it does not introduce any physical mechanism capable of making the disc react before the trapped population itself generates the corresponding potential perturbation.
    
    In any case, such asymmetries should be regarded as secondary effects. 
    The main result of the present analysis is that the self-gravity generated by the manifold-trapped population systematically precedes the subsequent large-scale spiral response of the disc. 
    If the observed structures were simply conventional density waves, there would be no obvious reason why a specific subset of particles should repeatedly and coherently generate deeper non-axisymmetric potential wells before the rest of the stellar population. 
    The fact that this behaviour emerges systematically within the manifold-trapped component is instead naturally interpreted as the dynamical signature of the action-reaction process underlying the delayed collective response of the disc, which we interpret as the defining characteristic of the \textit{material density waves} concept.

    \end{appendix}

\end{document}